# Strain enhancement of high-k dielectric response in (La/Sc)$_2$O$_3$ and LaScO$_3$: an ab-initio study.

Stas M. Avdoshenko and Alejandro Strachan[a]

*School of Materials Engineering and Birck Nanotechnology Center,*

*Purdue University, West Lafayette, Indiana, USA*

We use density functional theory within the generalized gradient approximation to characterize the dielectric response of rare earth oxides: (La,Sc)$_2$O$_3$ bixbyite, and LaScO$_3$ perovskite. We focus on the role of strain on the phonon contribution of the dielectric constant and find that, contrary to the classical expectation based on the Clausius-Mossotti relation, tensile volumetric strain and volume-conserving biaxial strain on the order of $\pm 1\%$ can lead to an increase in dielectric constant of up to 20%. The insight into the atomic mechanisms responsible for these effects and the quantitative results in this paper can contribute to the development and understanding of high-$\kappa$ materials.

---

[a]Electronic mail: strachan@purdue.edu

Miniaturization in the microelectronics industry has led to significant interest in materials with high dielectric constants (high-$\kappa$) to limit current leakage in complementary metal oxide semiconductor (CMOS) technology and[1,2] resulted in the incorporation of complex oxides like $HfO_2$ with $\kappa$=25 $\epsilon_o$[1]. Many promising high-$\kappa$ materials for such applications are rare earth oxides (REO) or REO-based ceramics that have been extensively studied in recent times[3-6,17]. The relatively large band gaps of these materials ~4.5-5 eV, a required property, limits the electronic contribution to the low frequency dielectric response; thus, their large $\kappa$, up to ~20-25 $\epsilon_o$[7,8,12,16], originates from the ionic or lattice components[3,11]. This same class of ceramics has also attracted attention for thermal barrier coatings[15] and in thermophotovoltaics[9]. In the case of thermal barrier coatings, their high emissivity at elevated temperatures can help remove energy from the structure radiatively. Thus, a fundamental understanding of the dielectric and optical properties of this class of materials and avenues to engineer their response is important from a basic science point of view but also can impact technological applications.

Significant efforts have been devoted to the development and understanding of high-$\kappa$ materials[11]. In the case of REOs prior studies involved exploring various chemical compositions and different allotropic phases[11] as well as strain[17]. In this Letter we use density functional theory (DFT) to explore strain as an avenue to control the low-frequency dielectric response of three REO's: $La_2O_3$, $Sc_2O_3$ and the perovskite $LaScO_3$. Recent experiments show that strain can cause significant changes in the dielectric response of REO's[17] but the mechanisms responsible for such effect are not well understood; furthermore, as our results show, the models and approximations typically used to interpret these results are not applicable in all cases.

The search for high-$\kappa$ materials is often guided by the classical Clausius-Mossotti formulation of dielectrics. According to this relation ($k = (3V_m + 8\pi\alpha^T)/(3V_m - 4\pi\alpha^T)$) the dielectric response of a homogeneous medium is governed by the interplay between total system polarizability ($\alpha^T$) and the molar volume ($V_m$)[10]. These two parameters can be modulated by an appropriate choice of RE element with different atomic radii and electronegativity in order to increase $\kappa$. In this continuum picture a smaller molar volume leads to increased permittivity if the local polarizability remains constant. Clausius-Mossotti can be used to assess the role of strain in dielectric response. For example, using this relationship Ref.[10] predicts the dielectric constant of hexagonal $La_2O_3$ to be 35 $\epsilon_0$ (an overestimation). More importantly, Clausius-Mossotti indicates a significant strain effect: assuming no change in polarizability a ± 3% volumetric strain would increase the dielectric constant to 56 $\epsilon_0$ (in compression) and reduce it to 27 $\epsilon_0$ (in tension). An atomic picture of dielectric response and our DFT results show different trends.

We use DFT to characterize the phonon and electronic contribution to dielectric response in the group of related materials cubic C-(La/Sc)$_2$O$_3$ and perovskite type LaScO$_3$. Our results show that both hydrostatic and volume conserving biaxial strains are able to enhance the dielectric response up to 20% from the unstrained value. The trends observed originate from the interplay between atomic polarization and normal modes of vibration that dominates dielectric response[12,18] and contradict Clausius-Mossotti.

*Systems.* The systems of interest are well-characterized high-$k$ materials, (La/Sc)$_2$O$_3$ and LaScO$_3$, and our goal is to explore whether strain can be used to significantly alter their dielectric properties. The complex bixbyite-type cubic structures Ln$_2$O$_3$ are related to the simpler fluorite(CaF$_2$) structure; doubling the cubic CaF$_2$ unit cell along each axis results in a structure almost identical to the C-type structure for C-Ln$_2$O$_3$. Under normal conditions, both La and Sc oxides crystallize into a body-centered cubic structure, the unit cell contains 3 wyckoff positions multiplied by I$\bar{a}$3 space symmetry group up to 40 in the primitive unit cell[24]. LaScO$_3$ crystallizes in a perovskite-like structure with P$nma$ symmetry[25].

*Methods.* All calculations were performed using the Quantum Espresso package[26] within the generalized gradient approximation (GGA) for the exchange and correlation potentials[27] and using norm-conserving pseudopotentials[28] with scalar relativistic corrections. We explicitly consider 11 electrons for La [5s$^2$5p$^6$5d$^2$6s$^{1.5}$6p$^{0.5}$] and Sc [3s$^2$3p$^6$4s$^2$3d$^1$] atoms with their respective core (Pd and Ne) electrons treated with pseudopotentials. For oxygen we describe 6 electrons explicitly [2s$^2$2p$^4$]. We use a kinetic energy cutoff of 30 Ry and sampling in reciprocal space is done via a 2×2×2 grid in (La/Sc)$_2$O$_3$ and 4×4×4 in the case of LaScO$_3$. All the structures were optimized via Hellmann-Feynman forces[29] with a tolerance of 0.01 eV/Å and stresses relaxed to 0.5$kbar$. Describing $f$-electrons in lanthanides and actinides is challenging within DFT due to their high-degree of localization.[13] Fortunately, La$^{+3}$ has an electronic configuration of Xe with an empty $f$-shell hidden deeply in the conduction band; the negligible role of $f$-states in bonding and dielectric response makes DFT suitable to describe this heavy element.

The dielectric response of materials is a function of the energy of transition modes and the associated transition moments. In the case of phonons, transition energies or frequencies of vibrational states are determined by the eigenvalues of the dynamical matrix. The transition moments are calculated in terms of the Born effective charge tensors and displacements given by eigenvectors of dynamical matrix. The Born effective charge tensors ($Z^*_{i,j} \sim \delta \mathbf{F_i}/\delta \mathbf{E_j}$) describe the on-site response of the system to a finite electric field $E_j$ in direction $j$ via an induced force $F_i$ in direction $i$. Alternatively, this tensor can be thought of as the change in polarization due to the displacement of atoms from their equilibrium positions. Born effective charges can be computed from finite differences or using perturbation theory. We use the second approach for the calculations in this paper. The oscillator strength tensors for possible transitions are then obtained from the atomic Born effective charge tensors and mass-weighted eigenmodes as: $\Omega_{\alpha\beta}^n = \sum_i Z_{\alpha\gamma}^i \vec{e}_{i\gamma}^n / m_i^{0.5} \cdot \sum_i Z_{\beta\gamma}^i \vec{e}_{i\gamma}^n / m_i^{0.5}$. Fermi's golden rule can then be used to obtain the phonon contribution to the imaginary part of the dielectric constant. The *Kramers-Kroning* transformation gives the real part of the phonon contribution to the dielectric constant as:

$$\varepsilon'^{\alpha\beta}(\omega) = \frac{4\pi}{V} \sum_n \frac{\Omega_{\alpha\beta}^n}{\omega_n^2 - \omega^2},$$

where the sum runs over all phonons, $n$, $\omega_n$ represents the frequency of mode $n$ and $\Omega_{\alpha\beta}^n$ its transition moment. In the remainder of the paper we will focus on the static dielectric constant ($\omega$=0) and averaging over all field orientations $\langle \varepsilon^{\alpha\beta}(0) \rangle_{\alpha\beta}$ will yield a theoretical expectation for polycrystalline materials. The electronic contribution to the dielectric constant ($\varepsilon'^{el}$) is obtained from the relaxed electronic structure of the systems. The total static static dielectric constants (electronic plus ionic) obtained for the materials of interest are reported in Table I.

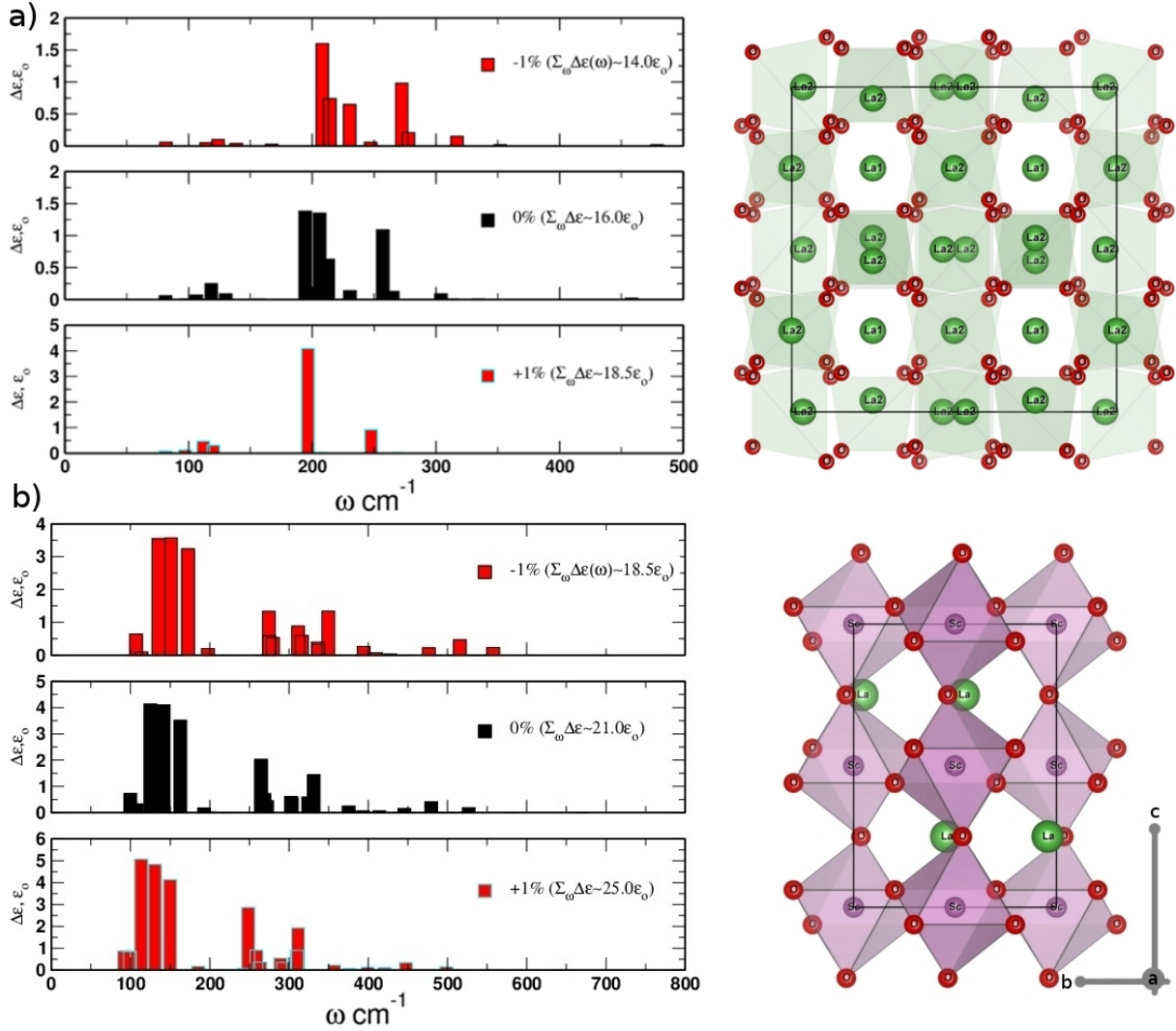

FIG. 1. a) An example of the mode-to-mode contribution for total dielectric response for unstressed $La_2O_3$ system (left). Atomistic structure of $La_2O_3$ with unit cell axis orientation (right). b) An example mode-to-mode contribution to a total dielectric response for unstressed $LaScO_3$ system (left). Atomistic structure of $LaScO_3$ with unit cell axis orientation (right).

*Results and discussion.*

For Ln=La/Sc our calculations predict equilibrium unit cell parameters of 11.32 and 9.94 Å respectively; these values are in good agreement with experimental observations 11.37(2)/9.8378(7)[24] and with previous theoretical predictions by *Delugas et al.*[16]. In the case of LaScO$_3$ our DFT-GGA lattice parameter predictions (a=5.71 Å, b= 5.59 Å, c=8.18 Å) are also in good agreement with experiments (a=5.6803(1) Å, b=5.7907(1) Å, c=8.0945(1) Å)[25]. We note that, as is often done in electronic structure calculations, the predicted lattice parameters correspond to zero temperature and ignore zero point energy[20]. Also, crystal symmetries were enforced during all structural relaxations.

Theoretical studies of the dielectric response of these materials under zero stress have been reported in recent years[12,16]. *Delugas et. al.* performed a detailed analysis of the contribution of individual modes to the dielectric response for various materials[12,16]. According to these reports DFT values for the phonon contributions to the dielectric constants for Ln2O3 system are 15.9 $\epsilon_o$ (for Ln=La) and 13.4 $\epsilon_o$ (for Ln=Sc); these are in good agreement with our predictions of 16.0 $\epsilon_o$ and 10.3 $\epsilon_o$ respectively (see Table I). The small discrepancy may originate from differences in the pseudopotential used for oxygen. These theoretical values are in rather good agreement with experimental data for Sc$_2$O$_3$ ($\epsilon$ = 13-14 $\epsilon_o$)[19] and La$_2$O$_3$ ($\epsilon$ = 18-27 $\epsilon_o$)[21]. The experimental value for LaScO$_3$ $\epsilon$ = 26 $\epsilon_o$[22] is also consistent with our predictions.

TABLE I. Summary of calculated phonon ($\epsilon^{ph}$) and electronic($\epsilon^{el}$) contributions and their sum in units of $\epsilon_o$ for each state (compression tension in %) for LaScO3. $\Delta E$ shows relative total energy in **eV** for each system.

| System | State | $\epsilon^{el}(0)$ | $\epsilon^{ph}(0)$ | $\epsilon^{el+ph}(0)$ | $\Delta$ E |
|---|---|---|---|---|---|
| **La$_2$O$_3$** | | | | | |
| | -1.0 | 4.2 | 14.0 | 18.2 | 0.24 |
| | 0.0 | 4.2 | 16.0 | 20.2 | 0.00 |
| | +1.0 | 4.2 | 18.5 | **22.7** | 0.15 |
| **Sc$_2$O$_3$** | | | | | |
| | -1.0 | 4.6 | 8.6 | 13.6 | 0.35 |
| | 0.0 | 4.6 | 10.3 | 14.9 | 0.00 |
| | +1.0 | 4.6 | 11.5 | **16.1** | 0.10 |
| **LaScO$_3$** | | | | | |
| | -1.0 | 4.8 | 18.5 | 23.3 | 0.60 |
| | 0.0 | 4.8 | 21.0 | 25.8 | 0.00 |
| | +1.0 | 4.8 | 25.0 | **29.8** | 0.20 |
| **LaScO$_3$** | | | | | |
| | -1.0$^{ca}$ | 4.8 | 19.6 | 24.4 | 0.10 |
| | -1.0$^{ba}$ | 4.8 | 22.5 | **27.4** | 0.44 |
| | 0.0 | 4.8 | 21.0 | 25.8 | 0.00 |
| | +1.0$^{ca}$ | 4.8 | 23.5 | **28.3** | 0.56 |
| | +1.0$^{ba}$ | 4.8 | 21.1 | 25.9 | 0.22 |

The ionic contribution to the static dielectric constant is given by the sum of volume-normalized, transition moments divided by frequency squared. This explains why the combination of highly-polarized Me-O bonds and low vibrational frequencies due to heavy atomic masses results in REOs being attractive high-$\kappa$ materials. Interestingly, this atomistic pic-ture of dielectric response also provides insight into a possible avenue to increase the dielectric constant of a material: decreasing its vibrational frequencies via strain. Tensile strain increases the periodicity of the potential energy landscape experienced by ions; the consequent reduction in curvature would shift the material vibrational spectra towards lower frequencies. If the strain does not significantly affect normal modes and Born effective charges, an increase in the dielectric constant would result. On the contrary, compression tends to squeeze potential wells causing a blueshift in the spectra and, potentially, a reduction in dielectric constant.

We performed DFT calculations of the dielectric response of all three materials for volumetric strain where $e_{xx}=e_{yy}=e_{zz}=\pm 1\%$. (We use $e$ for strain to avoid confusion with dielectric constant.) This corresponds to a volumetric strain (trace of the strain tensor for small deformations) of approximately 3% with predicted levels of stress in the range 5-10 GPa. For each strain, the atomic coordinates are fully relaxed and the electronic and ionic contributions to dielectric response are computed as described above. The results are shown in Table I; while the electronic contribution to the dielectric constant is not affected significantly by strain (<1%) the ionic contributions show the expected trends with tension leading to an increase in dielectric constant, up to 20% in the case of the perovskite. To understand the origin of such a significant effect, Figures 1(a,b) show the individual contribution of normal modes to the static dielectric constant ($\frac{4\pi}{V} \Omega_i/w_i^2$) for $La_2O_3$ [Fig. 1(a)] and $LaScO_3$ [Fig. 1(b)]. In each material, we show results for three levels of strains and the Supplementary Material[34] includes tables with spectral contributions for all cases. As expected, tension leads to softening of the modes in all three materials and this contributes to the increase in dielectric constant. In the case of $LaScO_3$, see Fig. 1(b), the spectral features of dielectric response exhibit little change with strain beyond the aforementioned softening; we find that 1% tensile strain reduces the average vibrational frequency by ∼5-8% while the Born effective charges remain relatively unchanged (in average, the effective charge on the metallic and oxygen atoms has been changed by ∼0.5-1.0%). This results in an increase of the ionic dielectric constant by 5 $\epsilon_o$ leading to a total value of ∼ 30 $\epsilon_o$ (see Supplementary Material[34]). The general trends are similar in the case of $La_2O_3$, Fig. 1(a); however, a single mode at approximately 200 cm$^{-1}$ contributes significantly in tension, while several different modes contribute in this frequency range for unstrained and compressive cases. The Supplementary Material[34] includes tables with the individual contributions of each vibrational mode.

Although, the magnitude of strain explored is achievable experimentally via epitaxial integration of appropriately chose materials, volumetric strain is only of academic interest as it is not experimentally accessible outside of laboratory setups like diamond anvil cells[30,31]. Thus, we now focus on volume conserving bi-axial strain with strain components: $e_{xx}=e_{yy}=e_0$ and $\epsilon_{zz}=1/(1+e_0)^2-1$. Of course, volume conserving bi-axial strain is also an idealized condition but it is useful to consider it as our two strain paths span the entire range of possible deformations: from pure volumetric strains with no deviatoric (shear) component to pure deviatoric deformation with no volume change. Experimentally realizable bi-axial and uniaxial stress are intermediate between these two extreme cases[32,33].

We focus the volume-conserving bi-axial study on $LaScO_3$ which exhibits the highest dielectric response and where symmetry operations are not affected by bi-axial strain. For this crystal structure, see Figure 1(c), axis $a$ and $b$ are structurally similar, thus we focus on two strain paths: bi-axial tension and compression in the (a-b) plane and bi-axial tension and compression on the (a-c) plane. The effect of bi-axial stress on dielectric response is shown at the bottom of Table I. We find that the level of dielectric response enhancement can be similar that what is achievable with volumetric deformation. Interestingly, our results show that tension along the $c$ direction is critical to increase dielectric response; note that 1% compression on the $a-b$ plane is accompanied by approximately 2% tension along $c$. Interestingly, we find that the level of enhancement in dielectric response achievable with 1% strain is similar to that observed in thin REO films.[17]

In summary, we used DFT calculations to characterize the role strain in dielectric response of REO's of interest in high-$\kappa$ applications. An atomic picture of dielectric response provides insight into avenues to enhance dielectric response that complements (and may contradict) the classical description based on Claussius-Mossotti. Our DFT results show that volumetric and biaxial strains on the order +1% can enhance dielectric response up to 20% with the enhancement dominated by softening of vibrational modes; these results indicate that experimental exploration along these directions would be important. These results and physical insight are critical to understand the dielectric enhancement in thin films[17].

*Acknowledgements.* We thank P. Delugas and V. Fiorentini for helpful discussions and S. Sullivan for reading this manuscript. This work supported by US Air Force Office of Scientific Research via Grant: FA9550-11-0079 (Program Manager: Ali Sayir).